# Quantum morphogenesis: A variation on Thom's catastrophe theory

Diederik Aerts[1], Marek Czachor[2], Liane Gabora[1], Maciej Kuna[2], Andrzej Posiewnik[3], Jarosław Pykacz[3], Monika Syty[2]

[1] Centrum Leo Apostel (CLEA) and Foundations of the Exact Sciences (FUND)
Brussels Free University, 1050 Brussels, Belgium
[2] Wydział Fizyki Technicznej i Matematyki Stosowanej
Politechnika Gdańska, 80-952 Gdańsk, Poland
[3] Wydział Fizyki i Matematyki, Uniwersytet Gdański, 80-952 Gdańsk, Poland

Non-commutative propositions are characteristic of both quantum and non-quantum (sociological, biological, psychological) situations. In a Hilbert space model states, understood as correlations between all the possible propositions, are represented by density matrices. If systems in question interact via feedback with environment their dynamics is nonlinear. Nonlinear evolutions of density matrices lead to phenomena of morphogenesis which may occur in non-commutative systems. Several explicit exactly solvable models are presented, including 'birth and death of an organism' and 'development of complementary properties'.

## I. INTRODUCTION

René Thom's catastrophe theory is an attempt of finding a universal mathematical treatment of morphogenesis [1] understood as a temporally stable change of form of a system. The theory works at a meta-level and does not crucialy depend on details of interactions that form a concrete ecosystem, organism, or society. In order to achieve this goal the analysis must deal with qualitative classes of objects and has to possess certain universality properties.

The purpose of the present work is similar. We define a system by an abstract space of states. The set of propositions which define properties of the system is in general non-Boolean. In particular, propositions corresponding to the same property may not be simultaneously measurable if considered at different times. Also at the same time there may exist sets of mutually inconsistent propositions.

Although formal logical systems of this type are well known from quantum mechanics [2] it is also known that the scope of applications of non-Boolean logic is much wider [3–7]. Practically any situation which involves *contexts* belongs to this cathegory. Formally a context means that a logical value associated with a given proposition depends on a history of the system. In particular, an order in which questions are asked is not irrelevant.

The systems we shall consider are probabilistic. The morphogenesis will be described in terms of probabilities or uncertainties associated with given sets of propositions. The contextual nature of the propositions will require a representation of probabilities different from the Kolmogorovian framework [8] of sets and commuting projectors (characteristic functions). Propositions will be represented by projectors on subspaces of a Hilbert space.

Another element which we regard as crucial is a *feedback*. Feedback means that the system under consideration interacts with some environment. The environment is influenced by the system and the system reacts to the changes of the environment. Even simplest models of such interactons lead effectively to nonlinear evolution equations [9]. Therefore, instead of modeling the interaction we will say that the feedback is present if the dynamics of the system is nonlinear, with some restrictions on the form of nonlinearity.

A system which interacts with environment is statistically characterized by nontrivial conditional probabilities. In the language of non-Kolmogorovian probability calculus this implies that states are not given by simple tensor products of states. On the other hand, a simple tensor describes a state involving no correlations and hence neither interactions nor feedback.

As a consequence, the nonlinearity representing feedback should disappear if the system in question and the environment are in a product state. The latter property may be used to reduce the class of admissible nonlinear evolutions. In the Hilbert-space language the state of a subsystem is represented by a statistical operator $\rho$ which is not a projector (i.e. $\rho^2 \neq \rho$) whenever the state of the composite system subsystem+environment is not a product state. Therefore, the condition $\rho^2 = \rho$ characterizes states of subsystems which do not interact with environments. This leads to the following restriction: The dynamics of $\rho$ is linear if $\rho^2 = \rho$.

The latter condition is still not restrictive enough since it can be satisfied by both dissipative and non-dissipative evolutions [10]. We shall restrict the dynamics to Hamiltonian systems. In the present paper the Hamiltonian functions will be time independent, which roughly means that the form of the feedback does not change in time.

Finally, we want to make the discussion *universal*. By this we mean two things: (1) The Hamiltonian functions should be typical of a very large class of dynamical systems, and (2) the results should not crucially depend on the form of a feedback, but more on the very fact that the feedback is present.

The most universal Hamiltonian functions seem to correspond to Hamiltonians with equally spaced spectra or, more precisely, whose spectra contain equally spaced subsets. The class of Hamiltonians includes harmonic oscillators, quantum fields, spin systems, ensembles of identical objects, and many others. Quite recently the role of Hamiltonians of the harmonic oscillator type was shown to be relevant to the dynamics of a stock market [11].

A linear Hamiltonian dynamics of $\rho$ is given by the von Neumann equation

$$i\dot\rho = \hat\omega\rho, \tag{1}$$

with $\hat\omega\rho = [H, \rho]$. The equation (1) may be also regarded as an abstract representation of a harmonic oscillator. An oscillator which occurs in many applications in biological sciences is however the nonlinear oscillator [12], whose abstract version reads

$$i\dot\rho = \sum_j \hat\omega_j f_j(\rho). \tag{2}$$

The 'generic' equation which is the basis of our analysis is therefore the von Neumann-type equation

$$i\dot\rho = \sum_j [H_j, f_j(\rho)]. \tag{3}$$

The index $j$ is responsible for the possibility of having different parts of the system which differently interact via the feedback. For the sake of simplicity in this paper we restrict the analysis to only one $H$ and a single $f$:

$$i\dot\rho = [H, f(\rho)]. \tag{4}$$

The only assumptions we make about $f$ is that this is a standard operator function in the sense accepted in spectral theory of self-adjoint operators, and that it should be linear whenever there is no feedback. Nontrivial example satisfying all the above requirements is an arbitrary polynomial

$$f(\rho) = a_0 + a_1\rho + \ldots + a_n\rho^n. \tag{5}$$

## II. RELATION TO REACTION-DIFFUSION MODELS

The typical reaction-diffusion models are of the form [13,14]

$$i\dot X = \hat\omega X + \hat\omega_1 f(X) \tag{6}$$

where $\hat\omega = A\nabla^2$, and $A$ and $\hat\omega_1$ are, in general complex, matrices and $X$, $f(X)$ are vectors. Particular cases of (6) are the Swift-Hohenberg, $\lambda - \omega$, and Ginzburg-Landau models [15–18].

To illustrate what kind of models we arrive at consider the quadratic nonlinearity $f(\rho) = \rho^2$ and the harmonic oscillator Hamiltonian $H = \sum_{n=0}^{\infty} n|n\rangle\langle n|$. In the simplest case of a one-dimensional harmonic oscillator our nonlinear von Neumann equation $i\dot\rho = [H, \rho^2]$ reads in position space

$$i\dot\rho(x,y) = \big(-\partial_x^2 + \partial_y^2 + x^2 - y^2\big) \int dz\, \rho(x,z)\rho(z,y). \tag{7}$$

So even simplest cases lead to rather complicated integro-partial-differential nonlinear equations. The no-feedback condition implies $\rho(x,y) = \psi(x)\bar\psi(y)$, $\int dz\,\bar\psi(z)\psi(z) = 1$, and the equation can be separated yielding the Schrödiger equation

$$i\dot\psi(x) = (-\partial_x^2 + x^2 + \text{const})\psi(x). \tag{8}$$

The Schrödinger equation may be regarded as a diffusion equation in complex time. Similarly, the nonlinear von Neumann equations can be mapped into diffusion type equations by replacing $t$ by $it$, or by admitting non-Hermitian $H_j$. The Darboux techniques we are using are not restricted to Hermitian operators.

The self-switching solutions discussed below correspond to certain $\rho(x,y) \neq \psi(x)\bar\psi(y)$. The 'patterns' we find in explicit examples are illustrated by the probability densities

$$p_{t,x} = \langle x|\rho_t|x\rangle = \rho_t(x,x). \tag{9}$$

## III. ENTITIES IN ENVIRONMENTS

Consider two Hilbert spaces: $\mathcal{H}_E$ describing an 'environment' and spanned by vectors $|E\rangle$, and $\mathcal{H}_e$ describing an 'entity' and spanned by vectors $|e\rangle$. The composite system 'environment+entity' is represented by either a state vector

$$|\Psi\rangle = \sum_{E,e} \Psi_{Ee}|E,e\rangle = \sum_{E,e} \Psi_{Ee}|E\rangle \otimes |e\rangle \qquad (10)$$

or by a density matrix

$$\rho = \sum_{EE'ee'} \rho_{EE'ee'}|E,e\rangle\langle E',e'| \qquad (11)$$

Assuming that all expectation values of random variables are represented in terms of quantum averages we can write

$$\langle A \rangle_\Psi = \langle \Psi|A|\Psi\rangle \qquad (12)$$

or

$$\langle A \rangle_\rho = \operatorname{Tr} \rho A. \qquad (13)$$

Of particular interest are averages representing certain statistical quantities associated only with the entities, i.e. of the form

$$\langle I \otimes A_e \rangle_\Psi = \langle \Psi|I \otimes A_e|\Psi\rangle = \operatorname{Tr}_e \rho_e A_e \qquad (14)$$

or

$$\langle I \otimes A_e \rangle_\rho = \operatorname{Tr} \rho(I \otimes A_e) = \operatorname{Tr}_e \rho_e A_e. \qquad (15)$$

The reduced density matrices $\rho_e$ are defined, respectively, by

$$\rho_e = \operatorname{Tr}_E |\Psi\rangle\langle\Psi| = \sum_{Eee'} \Psi^*_{Ee}\Psi_{Ee'}|e\rangle\langle e'| \qquad (16)$$

or

$$\rho_e = \operatorname{Tr}_E \rho = \sum_{Eee'} \rho_{EEee'}|e\rangle\langle e'| \qquad (17)$$

In particular, for *product states*, i.e. those of the form

$$|\Psi\rangle = \sum_{E,e} \psi_E \phi_e |E,e\rangle = |\psi\rangle \otimes |\phi\rangle \qquad (18)$$

or

$$\rho = \sum_{EE'ee'} \varrho_{EE'}\sigma_{ee'}|E,e\rangle\langle E',e'| = \varrho \otimes \sigma \qquad (19)$$

the reduced density matrices are, respectively,

$$\rho_e = \operatorname{Tr}_E|\Psi\rangle\langle\Psi| = |\phi\rangle\langle\phi| \qquad (20)$$

and

$$\rho_e = \operatorname{Tr}_E \rho = \sigma. \qquad (21)$$

In such a case we say that the entity is uncorrelated with the environment, i.e. probabilities of events associated with the entity are independent of all the events associated with the environment.

States of composite systems are of a product form *if and only if* entities are uncorrelated with environments. Interactions of entities with environments destroy the product forms and introduce correlations.

Reduced density matrices corresponding to nontrivial correlations satisfy the condition

$$\rho_e^2 \neq \rho_e. \tag{22}$$

Any density matrix $\rho$ is Hermitian and positive. From the spectral theorem it follows that there exists a basis such that $\rho$ is diagonal. For example, any density matrix of an entity can be written in some basis as

$$\rho_e = \sum_e p_e |e\rangle\langle e| \tag{23}$$

Now consider a vector

$$|\Psi\rangle = \sum_e \sqrt{p_e} |\Psi_e\rangle \otimes |e\rangle \tag{24}$$

where $|\Psi_e\rangle \in \mathcal{H}_E$ are any orthonormal vectors belonging to the Hilbert space of the environment and $|e\rangle$ are the eigenvectors of $\rho_e$. Then

$$\text{Tr}_E |\Psi\rangle\langle\Psi| = \sum_e p_e |e\rangle\langle e|. \tag{25}$$

In other words, for any density matrix $\rho_e$ one can find a state of the composite system guaranteeing that its reduced density matrix is identical to $\rho_e$. In what follows we shall therefore assume that a given initial $\rho_e$ is a result of correlations of the entity with the environment. If $\rho_e^2 \neq \rho_e$ then the correlations are nontrivial.

## IV. FEEDBACK WITH THE ENVIRONMENT

Typical systems discussed in the biophysics literature involve nonlinearities given by non-polynomial functions $f$. One often encounters Hill and other functions which are continuous approximations to step functions. A simple one-dimensional reaction-diffusion model describing experiments on regeneration and transplantation in hydra involves nonlinearities with positive and negative powers [19]. The environment is here modelled by two densities describing concentration of activator and inhibitor producing cells. Essential to the model is the symmetry breaking of the two densities, a fact accounting for the nonsymmetric development of hydra. More refined models [20] do not need externally imposed inhomogeneities but involve environments acting as active chemicals. The aim of complicated feedback behaviors is to account for the observed symmetry breaking of the development of hydra without a need of putting the nonsymmetric elements by hand.

A close quantum analogue of biophysical dynamical systems is a 'general' nonlinear von Neumann equation (4) [21,22]. If $f$ is to represent a feedback, the nonlinear effect should disappear if the entity is uncorrelated with the environment. Assuming the whole system is represented by a state vector $|\Psi\rangle$, the lack of correlations implies that $|\Psi\rangle = |\psi\rangle \otimes |\phi\rangle$ and $\rho_e = |\phi\rangle\langle\phi|$. Such a reduced density matrix satisfies $\rho_e^2 = \rho_e$. The condition 'no correlations, no feedback' is formally translated into

$$[H, f(\rho)] = [H, \rho] \quad \text{if} \quad \rho^2 = \rho. \tag{26}$$

Let us note that the above restriction means that $\rho_e = |\phi\rangle\langle\phi|$ satisfies an equation which is equivalent to

$$i|\dot\phi\rangle = H|\phi\rangle. \tag{27}$$

The latter is a general linear Schrödinger equation. In the absence of feedback the entity evolves according to the rules of quantum mechanics, an assumption which is rather general and weak.

This property has also another interpretation which is entirely 'classical'. Consider a system consisting of $N$ classical harmonic oscillators with frequencies $\omega_1, \ldots \omega_N$. Denote by $H$ the diagonal matrix $\text{diag}(\omega_1, \ldots, \omega_N)$ and by $|\phi\rangle$ a column vector with entries $\phi_k = q_k + ip_k$. Then (27) is equivalent to the system of classical equations $\dot q_k = \omega_k p_k$, $\dot p_k = -\omega_k q_k$. As a consequence, the description we propose may be extended even to fully classical systems which are modeled by ensembles of oscillators which evolve linearly and independently in the absence of a feedback.

Now, what are the restrictions imposed on $f$ by (26)? As we have said before, a general density matrix has a form $\rho_e = \sum_e p_e |e\rangle\langle e|$, where $p_e$ are probabilities. The spectral theorem implies that

$$f(\rho_e) = \sum_e f(p_e)|e\rangle\langle e|. \tag{28}$$

The condition $\rho_e^2 = \rho_e$ implies that $p_e^2 = p_e$ whose solutions are 0 and 1. Therefore (26) is satisfied by any $f$ which fulfills $f(0) = 0$ and $f(1) = 1$. In practical computations one can relax (26) by requiring only

$$[H, f(\rho)] \sim [H, \rho] \quad \text{if} \quad \rho^2 = \rho \tag{29}$$

since having (29) one can always reparametrize the time variable $t$ so that (26) is satisfied. The polynomial mentioned in the Introduction belongs to this cathegory.

The equation (4) possesses a number of interesting general properties. For example, the quantities

$$h = \operatorname{Tr} H f(\rho) \tag{30}$$
$$c_n = \operatorname{Tr}(\rho^n), \tag{31}$$

for all natural $n$, are time independent. $h$ is the Hamiltonian function for the dynamics and, hence, plays the role of the average energy of the entity (the feedback energy included). An analogous situation occurs in nonextensive statistics where $h$ has an interpretation of internal energy [22,36]. A system with conserved $h$ is *closed*.

Conservation of $c_n$ implies that eigenvalues of $\rho$ are conserved. The latter property means that there are certain features of the system that occur with time independent probabilities. However, and this is very important, the features themselves change in time in a way which is rather unusual in physical systems and has many analogies in evolution of biological systems.

## V. SOLITON MORPHOGENESIS

There exists a class of solutions of (4) which exhibits a kind of a three-regime switching effect [23–25]: For times $-\infty < t \ll t_1$ the dynamics looks as if there was not feedback, then in the switching regime $t_1 < t < t_2$ a 'sudden' transition occurs which drives the system into a *new* state which for times $t_2 \ll t < \infty$ evolves again as if there was no feedback. Of course, the feedback is present for all times, but is 'visible' only during the switching period. Formally the effect is very similar to scattering between two asymptotically linear evolutions ('self-scattering'). One can additionally complicate the dynamics by introducing an external element which makes the form of the feedback time dependent. We shall illustrate the effect on explicit examples.

The general equation (4) belongs to the family of equations integrable by means of soliton methods. One begins with its Lax representation

$$z_\lambda \langle \psi | = \langle \psi | (\rho - \lambda H), \tag{32}$$
$$-i \langle \dot\psi | = \frac{1}{\lambda} \langle \psi | f(\rho). \tag{33}$$

The construction requires two additional Lax pairs

$$z_\nu \langle \chi | = \langle \chi | (\rho - \nu H), \tag{34}$$
$$-i \langle \dot\chi | = \frac{1}{\nu} \langle \chi | f(\rho), \tag{35}$$
$$z_\mu |\varphi\rangle = (\rho - \mu H)|\varphi\rangle, \tag{36}$$
$$i |\dot\varphi\rangle = \frac{1}{\mu} f(\rho)|\varphi\rangle. \tag{37}$$

The method of solving (4) is based on the following theorem establishing the Darboux covariance of the Lax pair (32), (33) [25].

**Theorem.** Assume $\langle \psi |$, $\langle \chi |$ and $|\varphi\rangle$ are solutions of (32), (33), (34)–(37) and $\langle \psi_1 |$, $\rho_1$, are defined by

$$\langle \psi_1 | = \langle \psi | \Big( \mathbf{1} + \frac{\nu - \mu}{\mu - \lambda} P \Big), \tag{38}$$
$$\rho_1 = \Big( \mathbf{1} + \frac{\mu - \nu}{\nu} P \Big) \rho \Big( \mathbf{1} + \frac{\nu - \mu}{\mu} P \Big), \tag{39}$$
$$P = \frac{|\varphi\rangle\langle\chi|}{\langle\chi|\varphi\rangle}. \tag{40}$$

Then

$$z_\lambda \langle \psi_1 | = \langle \psi_1 | (\rho_1 - \lambda H), \tag{41}$$

$$-i \langle \dot\psi_1 | = \frac{1}{\lambda} \langle \psi_1 | f(\rho_1), \tag{42}$$

$$i\dot\rho_1 = [H, f(\rho_1)]. \tag{43}$$

∎

Let us note that the theorem is valid even for non-Hermitian $H$, i.e. for systems which are open. However, in the present paper we restrict te analysis to closed (conservative) systems chracterized by self-adjoint $H$. Systems whose average population does not change belong to this class.

One of the strategies of finding the 'switching solutions' is the following. One begins with a seed solution $\rho$ such that the operator

$$\Delta_a := f(\rho) - a\rho, \tag{44}$$

where $[a, H] = [a, \rho] = 0$, satisfies $[\Delta_a, H] = 0$ and $\Delta_a$ is not a multiple of the identity. Now we can write

$$i\dot\rho = [H, f(\rho)] = a[H, \rho] \tag{45}$$

and

$$\rho(t) = e^{-iaHt} \rho(0) e^{iaHt}. \tag{46}$$

Taking the Lax pairs with $\mu = \bar\nu$ and repeating the construction from [23,24], we get

$$\rho_1(t) = e^{-iaHt}\Big(\rho(0) + (\bar\nu - \nu) F_a(t)^{-1} e^{-i\Delta_a t/\bar\nu}[|\chi(0)\rangle\langle\chi(0)|, H] e^{i\Delta_a t/\nu}\Big) e^{iaHt}, \tag{47}$$

where

$$F_a(t) = \langle \chi(0) | \exp\Big(i\frac{\bar\nu - \nu}{|\nu|^2} \Delta_a t\Big) | \chi(0) \rangle$$

and $\langle \chi(0)|$ is an initial condition for the solution of the Lax pair.

## VI. 'SUDDEN' MUTATION OF POPULATION

In our first example we consider the quadratic nonlinearity $f(\rho) = (1 - h)\rho + h\rho^2$. The parameter $h$ controls the strength of the feedback. However, for any $h$ and any density matrix satisfying $\rho^2 = \rho$ we find $f(\rho) = \rho$ and the feedback vanishes. This is consistent with our assumption that $\rho^2 = \rho$ characterizes systems which are not interacting with an environment. We take the Hamiltonian $H = \sum_{n=0}^{\infty} n|n\rangle\langle n|$ which may represent a system whose energy is proportional to the number of its elements. Solutions of the von Neumann equation are in general infinite-dimensional but in order to illustrate the morphogenesis we restrict the analysis to a finite dimension. The lowest dimension where the effect occurs is 3. Therefore we select a subspace spanned by three subsequent vectors $|k\rangle, |k+1\rangle, |k+2\rangle$. We will discuss a family, parametrized by $\alpha \in \mathbf{R}$, of self-switching solutions $\rho_t = \sum_{m,n=0}^{2} \rho_{mn} |k+m\rangle\langle k+n|$ of (4). The solution is completely characterized by the matrix of time-dependent coefficients $\rho_{mn}$. Here we only give the final result and postpone a detailed derivation to Sec. VIII where we analyze a generalization involving a greater number of 'different species'. The reader may check by a straightforward substitution that the matrix

$$\begin{pmatrix} \rho_{00} & \rho_{01} & \rho_{02} \\ \rho_{10} & \rho_{11} & \rho_{12} \\ \rho_{20} & \rho_{21} & \rho_{22} \end{pmatrix} = \frac{1}{15 + \sqrt{5}} \begin{pmatrix} 5 & \xi(t) & \zeta(t) \\ \bar\xi(t) & 5 + \sqrt{5} & \xi(t) \\ \bar\zeta(t) & \bar\xi(t) & 5 \end{pmatrix} \tag{48}$$

with

$$\xi(t) = \frac{(2 + 3i - \sqrt{5}i)\sqrt{3 + \sqrt{5}}\alpha}{\sqrt{3}(e^{\gamma t} + \alpha^2 e^{-\gamma t})} e^{i\omega_0 t}, \quad \zeta(t) = -\frac{9e^{2\gamma t} + (1 + 4\sqrt{5}i)\alpha^2}{3(e^{2\gamma t} + \alpha^2)} e^{2i\omega_0 t}$$

is indeed a solution of the von Neumann equation. The parameters are $\omega_0 = 1 - \frac{5+\sqrt{5}}{15+\sqrt{5}} h$, $\gamma = \frac{2}{15+\sqrt{5}} h$.

There exists a critical value $h_0 = \frac{15+\sqrt{5}}{5+\sqrt{5}}$ corresponding to $\omega_0 = 0$. Using the explicit position dependence of the eigenstates of the harmonic oscillator Hamiltonian we can make a plot illustrating the time dependence of the probability density $p_{t,x}$ in position space as a function of time and $h$.

The dynamics we encounter in this example is particularly suggestive for $h = h_0$ (Fig. 1) and resembles a mutation of the statistical ensemble described by $\rho$. The corresponding probability appears static for, roughly, $-\infty < t < -40$ and then also for $40 < t < \infty$. Switching is 'suddenly' triggered in a neighborhood of $t = 0$. Fig. 2 shows the evolution of the probability density at the origin $p_{t,0}$ as a function of time for different values of $h$. For $h \neq h_0$ the probability density is an oscillating function of time, but in the neighborhood of $t = 0$ one observes the 'mutation' which occurs for any $h \neq 0$, the longer the transition period the smaller $h$. Duration of the switching process is of the order $1/h$. For $h = 0$ the dynamics is linear (no feedback) and there is no switching. The example shows that there occurs a kind of uncertainty relation between the strength of the feedback and duration of the switching: The smaller the feedback the longer the switching period.

Let us note that the probability density shown at Fig. 1 has this particular shape since we have used the position-space wave functions characteristic of a quantum one dimensional harmonic oscillator (a Gaussian times Hermite polynomials). Had we chosen any other system which is isospectral to a one-dimensional harmonic oscillator (or any system with equally spaced spectrum, say, a 3D harmonic oscillator) we would have obtained a different shape of the probability density. Although different choices of $H$ imply different differential equations, their common feature is the effect of 'mutation'.

## VII. COMPOSITE ENTITIES: BIRTH AND DEATH OF AN ORGANISM

In this example we consider an *organism*, that is a composite entity which undergoes the feedback process as a whole. A simple model consists of a two-qubit system described by the Hamiltonian

$$H = H_1 \otimes \mathbf{1} + \mathbf{1} \otimes H_2. \tag{49}$$

The Hamiltonian does not contain an interaction term. However, the two subentities forming the 'organism' do not evolve independently. They are coupled to each other through the feedback with the envirinment, i.e. through the nonlinearity. As we shall see, they become asymptotically uncoulped at $t \to \pm\infty$. In a 'distant past' the system consists of uncorrelated subentities which, after a period of certain joint activity, become again uncorrelated in the future. An analogy with 'birth' and 'death' is striking, and justifies the name 'organism'.

To make the example concrete assume that

$$H = 2\sigma_x \otimes \mathbf{1} + \mathbf{1} \otimes \sigma_z \tag{50}$$

We will start with the non-normalized density matrix

$$\rho(0) = \frac{1}{2} \begin{pmatrix} 5+\sqrt{7} & 0 & 0 & 0 \\ 0 & 5-\sqrt{7} & 0 & 0 \\ 0 & 0 & 5+\sqrt{15} & 0 \\ 0 & 0 & 0 & 5-\sqrt{15} \end{pmatrix} \tag{51}$$

which is written in such a basis that

$$H = \begin{pmatrix} 1 & 2 & 0 & 0 \\ 2 & 1 & 0 & 0 \\ 0 & 0 & -1 & 2 \\ 0 & 0 & 2 & -1 \end{pmatrix}. \tag{52}$$

The density matrix

$$\rho(t) = \exp[-5iHt]\rho(0)\exp[5iHt] \tag{53}$$

is a solution of (4) with $f(\rho) = \rho^2$. Such a $\rho(t)$ describes simultaneously a dynamics of two non-interacting systems satisfying the linear von Neumann equation

$$i\dot\rho = 5[2\sigma_x \otimes \mathbf{1} + \mathbf{1} \otimes \sigma_z, \rho]. \tag{54}$$

To understand why this happenes it is sufficient to note that the solution satisfies

$$[H, \rho^2] = [H, 5\rho] = [5H, \rho]. \tag{55}$$

The environment does not trigger in this solution any switching, but only makes its evolution five times faster than in the absence of the feedback. The Darboux transformation when applied to $\rho(t)$ produces (for more details cf. [23]) the solution

$$\rho_1(t) = \exp[-5iHt]\rho_{\text{int}}(t)\exp[5iHt] \tag{56}$$

where

$$\rho_{\text{int}}(t) = \frac{1}{2}\begin{pmatrix} 5-\sqrt{7}\tanh 2t & 0 & \frac{-13i-3\sqrt{7}-\sqrt{15}-i\sqrt{105}}{8\cosh 2t} & \frac{-7i+3\sqrt{7}-3\sqrt{15}+i\sqrt{105}}{8\cosh 2t} \\ 0 & 5+\sqrt{7}\tanh 2t & \frac{15i+\sqrt{7}-\sqrt{15}-i\sqrt{105}}{8\cosh 2t} & \frac{\sqrt{7}+\sqrt{15}}{2\cosh 2t} \\ \frac{13i-3\sqrt{7}-\sqrt{15}+i\sqrt{105}}{8\cosh 2t} & \frac{-15i+\sqrt{7}-\sqrt{15}+i\sqrt{105}}{8\cosh 2t} & 5+\sqrt{15}\tanh 2t & 0 \\ \frac{7i+3\sqrt{7}-3\sqrt{15}-i\sqrt{105}}{8\cosh 2t} & \frac{\sqrt{7}+\sqrt{15}}{2\cosh 2t} & 0 & 5-\sqrt{15}\tanh 2t \end{pmatrix}. \tag{57}$$

Now the switching between the two asymptotic evolutions is triggered in the neighborhood of $t = 0$.

If we look at the subentities forming the organism we notice that they do not evolve independently. The easiest way of seeing this is to compute the reduced density matrices of the two subentities. Here we write explicitly the eigenvalues of the reduced density matrices. Both subsystems are two-dimensional so there are two eigenvalues for each reduced density matrix. They read

$$p_{\pm}(1) = \frac{1}{2} \pm \frac{\sqrt{15}-\sqrt{7}}{20}\tanh 2t, \quad \text{particle 1} \tag{58}$$

$$p_{\pm}(2) = \frac{1}{2} \pm \frac{\sqrt{26+2\sqrt{105}}}{40\cosh 2t}, \quad \text{particle 2}. \tag{59}$$

The asymptotics are

$$\rho_{\text{int}}(-\infty) = \frac{1}{2}\begin{pmatrix} 5-\sqrt{7} & 0 & 0 & 0 \\ 0 & 5+\sqrt{7} & 0 & 0 \\ 0 & 0 & 5-\sqrt{15} & 0 \\ 0 & 0 & 0 & 5+\sqrt{15} \end{pmatrix}, \tag{60}$$

$$\rho_{\text{int}}(+\infty) = \frac{1}{2}\begin{pmatrix} 5+\sqrt{7} & 0 & 0 & 0 \\ 0 & 5-\sqrt{7} & 0 & 0 \\ 0 & 0 & 5+\sqrt{15} & 0 \\ 0 & 0 & 0 & 5-\sqrt{15} \end{pmatrix} = \rho(0), \tag{61}$$

and therefore the dynamics represents asymptotically two non-interacting subentities. It is also interesting that the $+\infty$ asymptotics is $\rho_1(t) \approx \rho(t)$. At large times the 'organism' which 'dies' becomes practically indistinguishable from the one that never 'lived'.

The 'life' of the organism is the period of time when the two subentities exhibit certain joint activity. Computing the von Neumann entropies of reduced density matrices of the two subentities we can introduce a quantitative measure of this activity. The entropies of the two particles are shown in Fig. 3. The organism lives several units of time. Similar are the scales of time when the off diagonal matrix elements of $\rho_{\text{int}}(t)$ become non-negligible. It should be stressed that the entropy characterizing the entire organism is time independent (since eigenvalues of solutions of (4) are constants of motion for all $f$).

Although it is clear that the 'organism' behaves during the evolution as an indivisible entity, one should not confuse this indivisiblility with the so-called nonseparability discussed in quantum information theory. The organism we consider in the example is a two-qubit system and therefore one can check the separability of $\rho_1(t)$ by means of the Peres-Horodecki partial transposition criterion [26,27]: A two qubit density matrix $\rho$ is separable if and only if its partial transposition is positive. It turns out that partial transposition of $\rho_1(t)$ is positive for any $t$ and, hence, $\rho_1(t)$ is in this sense separable (has 'zero entanglement'). It is well known, however, that 'zero entanglement' does not mean 'no quantum correlations' in the system. The so called three-particle GHZ state [28] is fully entangled at the three-particle level in spite of the fact that all its two-particle subsystems are described by separable density matrices.

# VIII. MODEL WITH SEVERAL SPECIES

The models we have considered so far corresponded to a Hilbert space with basis vectors $|n\rangle$. The only characterization of a state was in terms of the quantum number $n$ which could be regarded as the number of elements of a given population. Now we want to extend the description to the situation where we have a population consisting of several species characterized by numbers $n_1, \ldots, n_N$. The basis vectors are

$$|\boldsymbol{n}\rangle = |n_1, \ldots, n_N\rangle = |n_1\rangle \otimes \ldots \otimes |n_N\rangle \tag{62}$$

and the Hamiltonian

$$H = \sum_{n_j} (n_1 + \ldots + n_N)|n_1, \ldots, n_N\rangle\langle n_1, \ldots, n_N| \tag{63}$$

$$= \sum_n E_n |\boldsymbol{n}\rangle\langle \boldsymbol{n}|. \tag{64}$$

The Hamiltonian has equally spaced spectrum and is formally very similar to those we have encountered in the previous sections. The difference is that now the energy eigenstates are highly degenerated, a property which is very useful from the perspective of constructing multi-parameter and higher-dimensional self-switching solutions.

For simplicity consider two species ($N=2$), the quadratic nonlinearity

$$f(\rho) = (1-h)\rho + h\rho^2,$$

and take some three energy eigenvalues $E_k$, $E_{k+m}$, $E_{k+2m}$. For each energy take $l+1$ vectors, which will be denoted by

$$|0_j\rangle = |k-j, j\rangle, \tag{65}$$
$$|1_j\rangle = |k+m-j, j\rangle, \tag{66}$$
$$|2_j\rangle = |k+2m-j, j\rangle, \tag{67}$$

$j = 0, 1, \ldots l \leq k$. We start with the unnormalized density matrix

$$\rho(0) = \sum_{j=0}^{l} \rho_j(0) \tag{68}$$

where

$$\rho_j(0) = \frac{a}{2}\Big(|0_j\rangle\langle 0_j| + |2_j\rangle\langle 2_j|\Big) + \frac{a + \sqrt{a^2 + 4(b-m^2)}}{2}|1_j\rangle\langle 1_j| - \frac{\sqrt{a^2 + 4b}}{2}\Big(|2_j\rangle\langle 0_j| + |0_j\rangle\langle 2_j|\Big). \tag{69}$$

Positivity of $\rho_j(0)$ restricts the parameters as follows: $0 < 4m^2 < a^2 + 4b < a^2$. The operator

$$\Delta_a = \rho(0)^2 - a\rho(0) = b\tilde{I} - m^2 \sum_{j=0}^{l} |1_j\rangle\langle 1_j| \tag{70}$$

commutes with $H$. We denote by $\tilde{I}$ and $\tilde{H}$ the restrictions of the identity $I$ and $H$ to the $3(l+1)$-dimensional subspace spanned by (65)–(67). We will write $\tilde{H} = \sum_{j=0}^{l} H_j$, where

$$H_j = \sum_{n=0}^{2} (k+nm)|n_j\rangle\langle n_j|. \tag{71}$$

Consider the eigenvalue problem

$$\big(\rho_j(0) - iH_j\big)|\varphi_j\rangle = z|\varphi_j\rangle. \tag{72}$$

We find that the two solutions

$$|\varphi_j^{(1)}\rangle = -\frac{2im + \sqrt{a^2 + 4(b - m^2)}}{\sqrt{2}\sqrt{a^2 + 4b}}|0_j\rangle + \frac{1}{\sqrt{2}}|2_j\rangle \tag{73}$$

$$|\varphi_j^{(2)}\rangle = |1_j\rangle \tag{74}$$

correspond to the same $j$-independent eigenvalue

$$z = \frac{a + \sqrt{a^2 + 4(b - m^2)}}{2} + (k + m)i \tag{75}$$

and therefore

$$\bigl(\rho(0) - iH\bigr)|\varphi\rangle = z|\varphi\rangle. \tag{76}$$

with the same $z$ for any

$$|\varphi\rangle = \sum_{j=0}^{l} \Bigl(\alpha_j|\varphi_j^{(1)}\rangle + \beta_j|\varphi_j^{(2)}\rangle\Bigr). \tag{77}$$

The self-switching solution can thus be constructed by means of $|\varphi\rangle$ and reads

$$\rho_1(t) = e^{-i(1+h(a-1))Ht}\Bigl(\rho(0) + 2iF_a(t)^{-1}e^{-h\Delta_a t}[|\varphi\rangle\langle\varphi|, H]e^{-h\Delta_a t}\Bigr)e^{i(1+h(a-1))Ht}, \tag{78}$$

with

$$F_a(t) = \langle\varphi|\exp\bigl(-2h\Delta_a t\bigr)|\varphi\rangle = e^{-2hbt}\sum_{j=0}^{l}\Bigl(|\alpha_j|^2 + e^{2hm^2 t}|\beta_j|^2\Bigr) = e^{-2hbt}\Bigl(|\alpha|^2 + e^{2hm^2 t}|\beta|^2\Bigr). \tag{79}$$

Probabilities analogous to Fig. 1 are found if $a$ and $h$ are tuned in a way which eliminates the oscillating part $e^{-i(1+h(a-1))Ht}$, i.e. for $h = 1/(1 - a)$. In this case

$$\rho_1(t) = \rho(0) + 2i\Bigl(|\alpha|^2 + e^{2m^2 t/(1-a)}|\beta|^2\Bigr)^{-1}\sum_{j,j'=0}^{l}\Bigl[\bigl(\alpha_j|\varphi_j^{(1)}\rangle + \beta_j e^{\frac{m^2 t}{1-a}}|\varphi_j^{(2)}\rangle\bigr)\bigl(\bar{\alpha}_{j'}\langle\varphi_{j'}^{(1)}| + \bar{\beta}_{j'} e^{\frac{m^2 t}{1-a}}\langle\varphi_{j'}^{(2)}|\bigr), H\Bigr] \tag{80}$$

The vectors

$$|\Phi_j(t)\rangle = \alpha_j|\varphi_j^{(1)}\rangle + \beta_j e^{\frac{m^2 t}{1-a}}|\varphi_j^{(2)}\rangle = |\phi_j(t)\rangle \otimes |j\rangle \tag{81}$$

where

$$|\phi_j(t)\rangle = \frac{\alpha_j}{\sqrt{2}}\Bigl(-\frac{2im + \sqrt{a^2 + 4(b - m^2)}}{\sqrt{a^2 + 4b}}|k - j\rangle + |k + 2m - j\rangle\Bigr) + \beta_j e^{\frac{m^2 t}{1-a}}|k + m - j\rangle \tag{82}$$

are orthogonal for different $j$. Denoting $H = H_1 \otimes I + I \otimes H_2$ one can easily compute the reduced density matrix of the first species,

$$\rho_1^I(t) = \text{Tr}_2\rho_1(t) = \rho^I(0) + 2i\Bigl(|\alpha|^2 + e^{2m^2 t/(1-a)}|\beta|^2\Bigr)^{-1}\Bigl[\sum_{j=0}^{l}|\phi_j(t)\rangle\langle\phi_j(t)|, H_1\Bigr]. \tag{83}$$

The entire information about the dynamics of the first species is encoded in $\rho_1^I(t)$. Changes of properties of the species are given by the matrix elements

$$\langle n|\rho_1^I(t)|n'\rangle = \langle n|\rho^I(0)|n'\rangle + 2i(n' - n)\frac{\sum_{j=0}^{l}\langle n|\phi_j(t)\rangle\langle\phi_j(t)|n'\rangle}{|\alpha|^2 + e^{2m^2 t/(1-a)}|\beta|^2}. \tag{84}$$

An immediate conclusion from the above formula is that for $n = n'$ the expression is time independent. It follows that the number of elements of the ensemble does not change during the evolution. What changes are certain properties of the ensemble.

**A. Example: $k = m = 1$, $j = 0, 1$, $a = 5$, $b = -4$**

The two-species states in the subspace in question are

$$|0_0\rangle = |1, 0\rangle, \tag{85}$$
$$|1_0\rangle = |2, 0\rangle, \tag{86}$$
$$|2_0\rangle = |3, 0\rangle, \tag{87}$$
$$|0_1\rangle = |0, 1\rangle, \tag{88}$$
$$|1_1\rangle = |1, 1\rangle, \tag{89}$$
$$|2_1\rangle = |2, 1\rangle. \tag{90}$$

The two-species initial seed density matrix is given by

$$\rho_0(0) = \frac{5}{2}\Big(|1,0\rangle\langle 1,0| + |3,0\rangle\langle 3,0|\Big) + \frac{5+\sqrt{5}}{2}|2,0\rangle\langle 2,0| - \frac{3}{2}\Big(|3,0\rangle\langle 1,0| + |1,0\rangle\langle 3,0|\Big), \tag{91}$$

$$\rho_1(0) = \frac{5}{2}\Big(|0,1\rangle\langle 0,1| + |2,1\rangle\langle 2,1|\Big) + \frac{5+\sqrt{5}}{2}|1,1\rangle\langle 1,1| - \frac{3}{2}\Big(|2,1\rangle\langle 0,1| + |0,1\rangle\langle 2,1|\Big), \tag{92}$$

$$\rho(0) = \rho_0(0) + \rho_1(0). \tag{93}$$

Assume $\alpha_0 = \alpha_1 = 1/\sqrt{2}$, $\beta_0 = e^{t_0/4}$, $\beta_1 = e^{t_1/4}$. Then

$$|\phi_0(t)\rangle = \frac{1}{2}\Big(-\frac{2i+\sqrt{5}}{3}|1\rangle + |3\rangle\Big) + e^{(t_0-t)/4}|2\rangle, \tag{94}$$

$$|\phi_1(t)\rangle = \frac{1}{2}\Big(-\frac{2i+\sqrt{5}}{3}|0\rangle + |2\rangle\Big) + e^{(t_1-t)/4}|1\rangle \tag{95}$$

Writing the restriction of $H$ to the 6-dimensional subspace as

$$\tilde{H} = \begin{pmatrix} 1 & 0 & 0 & 0 & 0 & 0 \\ 0 & 1 & 0 & 0 & 0 & 0 \\ 0 & 0 & 2 & 0 & 0 & 0 \\ 0 & 0 & 0 & 2 & 0 & 0 \\ 0 & 0 & 0 & 0 & 3 & 0 \\ 0 & 0 & 0 & 0 & 0 & 3 \end{pmatrix} \tag{96}$$

we can represent the two-species density matrix $\rho_1$ in the form

$$\rho_1 = \begin{pmatrix} \frac{5}{2}\mathbf{1} & \frac{2-i\sqrt{5}}{3}\xi & -\frac{3}{2}\mathbf{1} + \frac{2-i\sqrt{5}}{3}\zeta \\ \frac{2+i\sqrt{5}}{3}\xi^T & \frac{5+\sqrt{5}}{3}\mathbf{1} & i\xi^T \\ -\frac{3}{2}\mathbf{1} + \frac{2+i\sqrt{5}}{3}\zeta & -i\xi & \frac{5}{2}\mathbf{1} \end{pmatrix} \tag{97}$$

where $\mathbf{1}$ is the $2 \times 2$ unit matrix, $^T$ denotes transposition, and

$$\xi(t) = \frac{e^{t/4}}{e^{t/2} + e^{t_0/2} + e^{t_1/2}} \begin{pmatrix} e^{t_1/4} & e^{t_0/4} \\ e^{t_1/4} & e^{t_0/4} \end{pmatrix} \tag{98}$$

$$\zeta(t) = \frac{e^{t/2}}{e^{t/2} + e^{t_0/2} + e^{t_1/2}} \begin{pmatrix} 1 & 1 \\ 1 & 1 \end{pmatrix}. \tag{99}$$

One can verify by a straightforward calculation that ($h = -\frac{1}{4}$)

$$i\dot\rho_1 = \tfrac{5}{4}[H, \rho_1] - \tfrac{1}{4}[H, \rho_1^2]. \tag{100}$$

To illustrate the time variation of statistical quantities associated with the two-species system it is sufficient to visualize the behavior of matrix elements of $\rho_1$. There are only three types of functions occuring in $\rho_1$:

$$F(t) = \frac{e^{t/2}}{e^{t/2} + e^{t_0/2} + e^{t_1/2}}, \tag{101}$$

$$F_0(t) = \frac{e^{(t+t_0)/4}}{e^{t/2} + e^{t_0/2} + e^{t_1/2}}, \tag{102}$$

$$F_1(t) = \frac{e^{(t+t_1)/4}}{e^{t/2} + e^{t_0/2} + e^{t_1/2}}. \tag{103}$$

The two parameters $t_0$, $t_1$, control two types of three-regime behaviors of $\rho_1$. The function $F$ is responsible for asymptotic properties of $\rho_1$ (via $\zeta$). Functions $F_0$, $F_1$ determine properties of the switching regime (via $\xi$). For $t_0 \ll t_1$ one finds $F_0(t) \approx 0$ for all $t$ and the switching is controlled by $F(t)$ and $F_1(t)$; the 'moment' of switching is shifted proportionally to $t_1$. For $t_0 \gg t_1$ one finds $F_1(t) \approx 0$ for all $t$ and the switching is controlled by $F(t)$ and $F_0(t)$; the 'moment' of switching does not depend on $t_1$ and is determined by $t_0$. Therefore the two types of switches are characterized by vanishing of those matrix elements of $\rho_1$ which contain either $F_0$ or $F_1$.

The asymptotic behavior of the system is given by

$$F_0(\pm\infty) = F_1(\pm\infty) = F(-\infty) = 0, \tag{104}$$
$$F(+\infty) = 1. \tag{105}$$

The reduced density matrices of single species are

$$\rho_1^I = \begin{pmatrix} \frac{5}{2} & \frac{2-i\sqrt{5}}{3}F_1 & -\frac{3}{2} + \frac{2-i\sqrt{5}}{3}F & 0 \\ \frac{2+i\sqrt{5}}{3}F_1 & 5 + \frac{\sqrt{5}}{2} & \frac{2-i\sqrt{5}}{3}F_0 + iF_1 & -\frac{3}{2} + \frac{2-i\sqrt{5}}{3}F \\ -\frac{3}{2} + \frac{2+i\sqrt{5}}{3}F & \frac{2+i\sqrt{5}}{3}F_0 - iF_1 & 5 + \frac{\sqrt{5}}{2} & iF_0 \\ 0 & -\frac{3}{2} + \frac{2+i\sqrt{5}}{3}F & -iF_0 & \frac{5}{2} \end{pmatrix} \tag{106}$$

$$\rho_1^{II} = \begin{pmatrix} \frac{15+\sqrt{5}}{2} & \frac{2-i\sqrt{5}}{3}F_1 + iF_0 \\ \frac{2+i\sqrt{5}}{3}F_1 - iF_0 & \frac{15+\sqrt{5}}{2} \end{pmatrix}. \tag{107}$$

Of course, all the density matrices are not normalized so that averages must be computed according to $\langle A \rangle = \operatorname{Tr} A\rho_1 / \operatorname{Tr} \rho_1$, etc. (note that $\operatorname{Tr} \rho_1$ is time independent).

## IX. MORPHOGENESIS OF COMPLEMENTARITY

According to the 'SSC theorem' [30,31] a density matrix $\rho$ is uniquely determined by correlations between all the possible propositions associated with a given system. Each matrix element of a $\rho$ can be given an interpretation in terms of probabilities associated with some proposition. In the Hilbert space language a proposition is a projector, i.e. an operator with eigenvalues 1 and 0 (logical 'true' and 'false'). Propositions which can be asked simultaneously are represented by commuting projectors. Propositions $P_1$, $P_2$ which do not commute are related by an uncertainty relation: The more is known about $P_1$ the less is known about $P_2$, and vice versa.

It is obvious that the above structures do not have to be associated with quantum systems. Just to give an example, many psychological tests are based on questionnaires which involve the same question asked many times in different contexts. The questions commute if the answer to a given question is always the same. However, in typical situations the same question has different answers within a single questionnaire. An ideal questionnaire involves all the possible questions asked in all the possible orders. In the Hilbert space formalism, where the questions are represented by projectors, an ideal questionnaire encodes all the possible correlations and thus, via the SSC theorem, is equivalent to a density matrix.

It is also known that there exist simple examples of systems whose logic is non-Boolean, but which do not allow a Hilbert space formulation [29]. The density matrix language will probably not suffice here and one has to admit a possibility of other state spaces and other nonlinear evolutions. The richness of available structures is immense.

Let us finally give examples of propositions whose averages (i.e. probabilities) change in time according to selected matrix elements of the self-switching solutions. The function $F(t)$ shown in Fig. 4 is associated with the proposition

$$P = \frac{1}{2} \begin{pmatrix} 1 & 0 & 1 & 0 \\ 0 & 0 & 0 & 0 \\ 1 & 0 & 1 & 0 \\ 0 & 0 & 0 & 0 \end{pmatrix} \tag{108}$$

corresponding to the first species as follows

$$p(t) = \frac{\text{Tr}\, P\rho_1^I(t)}{\text{Tr}\, \rho_1^I(t)} = \frac{1}{4}\frac{9+\sqrt{5}+8F(t)/3}{15+\sqrt{5}}. \tag{109}$$

Here $p(t)$ is the probability of the answer 'true' associated with $P$. Analogously $F_1(t)$ shown at Fig. 5 is associated with the proposition

$$P_1 = \frac{1}{2}\begin{pmatrix} 1 & 1 & 0 & 0 \\ 1 & 1 & 0 & 0 \\ 0 & 0 & 0 & 0 \\ 0 & 0 & 0 & 0 \end{pmatrix} \tag{110}$$

by means of

$$p_1(t) = \frac{\text{Tr}\, P_1\rho_1^I(t)}{\text{Tr}\, \rho_1^I(t)} = \frac{1}{4}\frac{15+\sqrt{5}+8F_1(t)/3}{15+\sqrt{5}}. \tag{111}$$

The evolution of the probabilities resembles the well known evolutions typically modelled by Hill functions [12] in the so-called sigmoidal response models [32–35,37]. Square deviations associated with the two propositions satisfy the uncertainty relation

$$\Delta P \Delta P_1 \geq \frac{1}{2}\left|\frac{\text{Tr}\,[P,P_1]\rho_1^I(t)}{\text{Tr}\,\rho_1^I(t)}\right| = \left|\frac{\sqrt{5}(e^{t/2}+e^{(t+t_0)/4})-(3+\sqrt{5})e^{(t+t_1)/4}}{12(15+\sqrt{5})(e^{t/2}+e^{t_0/2}+e^{t_1/2})}\right|. \tag{112}$$

For any fixed $t_0$, $t_1$, the right-hand-side of the inequality vanishes for $t \to -\infty$ and approaches $\sqrt{5}/[12(15+\sqrt{5})]$ for $t \to +\infty$. Fig. 6 shows this function for $t_1 = 0$. The two propositions which were not complementary in the past evolve into propositions satisfying an uncertainty relation.

In application to psychology a density matrix may represent an ideal questionnaire and, hence, a state of personality of a given individual. The morphogenesis we have discussed is a simple model of development of two complementary concepts. The model is simplified and perhaps too far fetched. However, philosophically this is not very far from the approaches of Thom [1] and particularly of Zeeman [38] in their catastrophe theory models of brain. More interesting in this context may be infinite dimensional cases whose preliminary analysis in terms of Darboux transformations for arbitrary $f(\rho)$ can be found in [25].

## X. DISCUSSION

The model of we have described satisfies the assumptions imposed by Thom on a *system of forms in evolution* ( [1] Chapter 1.2.A). The model is continuous and the morphogenesis is a result of soliton dynamics. In this respect the construction is analogous to nonlinear sigmoidal response models used in biochemistry [37]. What makes our construction essentially different from the models one finds in the literature is the role of non-commutativity of the system of propositions.

Non-commutative propositions are related by uncertainty principles and are typical of systems which cannot, without an essential destruction, be separated into independent parts. The examples can be taken not only from quantum physics, but also from sociology (communities), psychology (personalities), or biology (organisms). In all these cases the dynamics of a system consists of two parts: One generated by internal interactions, and the other corresponding to couplings with environment. We have considered only the simplest case where the internal dynamics is given a priori by a Hamiltonian of a harmonic oscillator type, and different parts of an organism (community, etc.) are coupled to each other only via the environment.

The coupling with environment leads to a feedback and, hence, nonlinear evolution. The systems we consider are conservative, but without difficulties can be generalized to explicitly time-dependent environments or non-Hermition Hamiltonians.

We model propositions by projectors on subspaces of a Hilbert space. States of the systems are represented by all the possible correlations between all the possible, even non-commuting, propositions. The choice of the Hilbert space language leads us therefore to a density matrix representation of states, and the dynamics is given in terms of nonlinear von Neumann equations. The formalism allows to consider morphogenesis of a completely new type, for example a development of complementary properties.

The class of solutions which has an interpretation in terms of morphogenesis has features which do not crucially depend on the form of the nonlinearity but more on the very presence of a feedback. The exact time development (say, duration) of morphogenesis does depend on initial conditions or the form of nonlinearity. However, the modifications do not influence the asymptotics, which is the qualitative element of the dynamics. For example, an organism which was 'born' has to 'die' but when and how will this occur depends on many details which are qualitatively irrelevant.

Instead of conclusions let us quote René Thom's final remarks from his early work on topological models in biology [39]:

"Practically any morphology can be given such a dynamical interpretation, and the choice between possible models may be done, frequently, only by qualitative appreciacion and mathematical sense of elegance and economy. Here we do not deal with a scientific theory, but more precisely with a *method*. And this method does not lead to specific techniques, but, strictly speaking, to *an art of models*. What may be, in that case, the ultimate motivation to build such models? They satisfy, I believe, a very fundamental epistemological need... If scientific progress is to be achieved by other means than pure chance and lucky guess, it relies necessarily on a *qualitative understanding* of the process studied. Our dynamical schemes (...) provide us with a very powerful tool to reconstruct the dynamical origin of any morphological process. They will help us, I hope, to a better understanding of the structure of many phenomena of animate and inamite nature, and also, I believe, of our own structure".

## ACKNOWLEDGMENTS

The work of MC and MK is a part of the KBN project 5 P03B 040 20. We would like to acknowledge the support of the Flemish Fund for Scientific Research (FWO project G.0335.02).

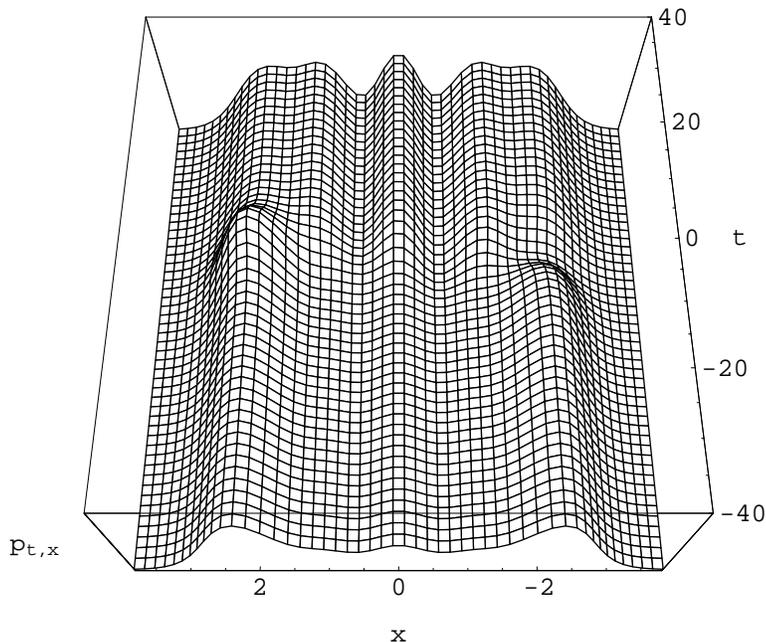

FIG. 1. The the probability density $p_{t,x} = \langle x|\rho_t|x\rangle$ for the critical value $h_0 = (15 + 5^{1/2})/(5 + 5^{1/2})$ as a function of time and $x$ for $-40 < t < 40$ (in arbitrary units). The three regimes are clearly visible. The probability interpolates between asymptotic probabilities which are constant in time. The visible switching (morphogenesis) begins around $t = -30$ and takes approximately 30 units of time. For later times the probability density becomes indistinguishable from the new asymptotic state.

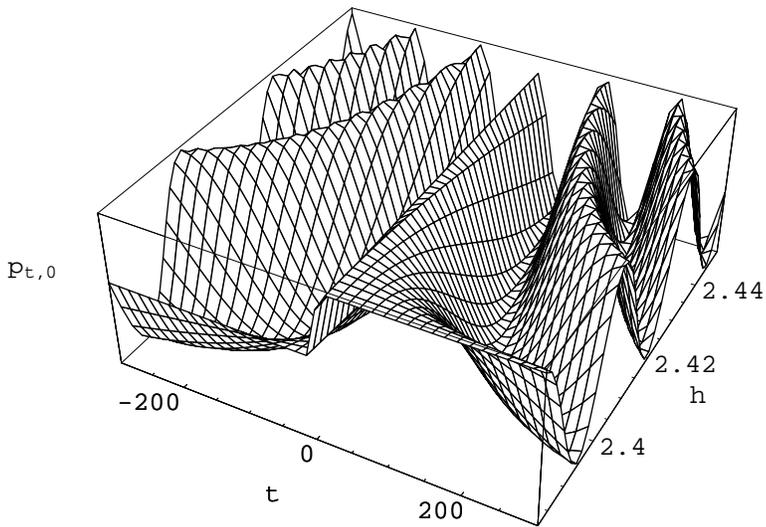
FIG. 2. Probability density $p_{t,0}$ at $x = 0$ as a function of time and $h$ for $h_0 \leq h \leq 2.45$. For $h > h_0$ the switching around $t = 0$ takes place between two different asymptotic oscillating probability densities. The switching is absent only for $h = 0$ (not shown) where the dynamics is linear.

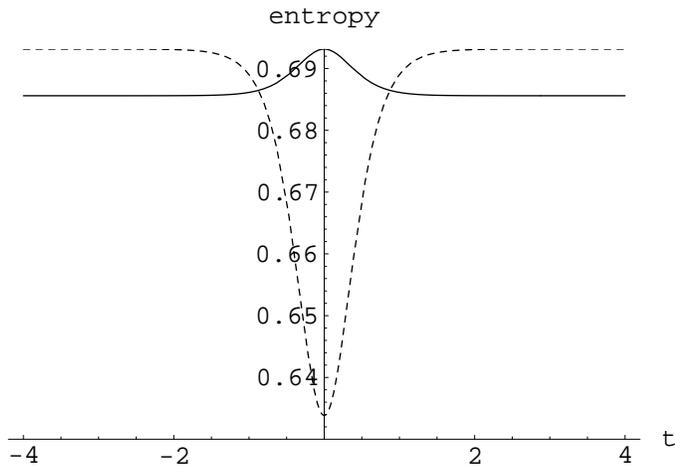
FIG. 3. Life and death of the two-qubit organism: the von Neumann entropies of particles 1 (solid) and 2 (dashed). The times where the particles are practically independent correspond to the flat parts of the plots.

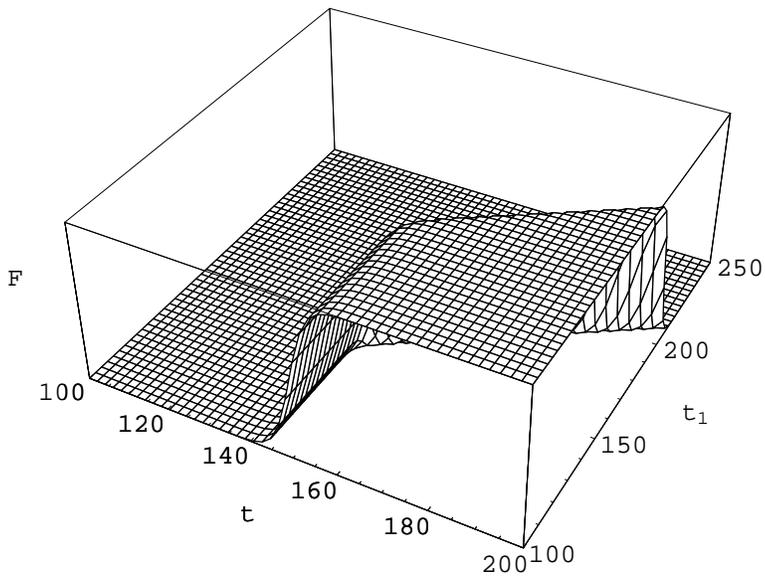

FIG. 4. $F(t)$ as a function of $t$ and $t_1$ for $t_0 = 150$.

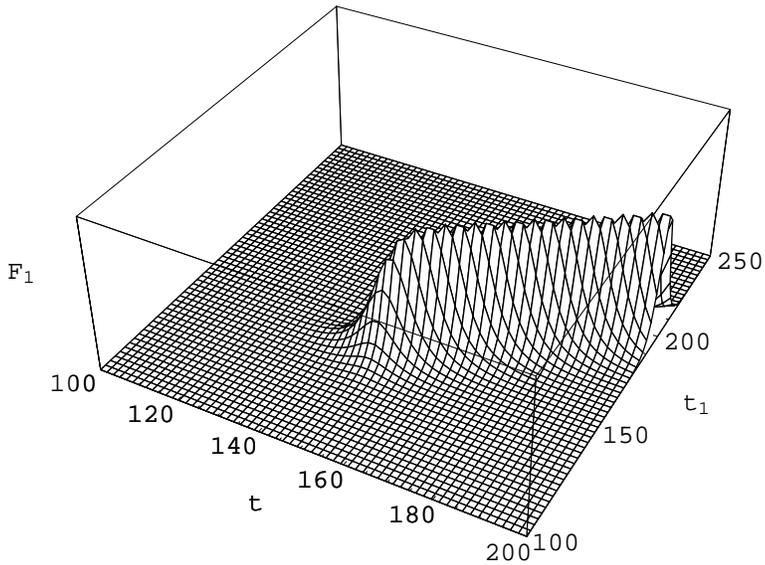

FIG. 5. $F_1(t)$ as a function of $t$ and $t_1$ for $t_0 = 150$.

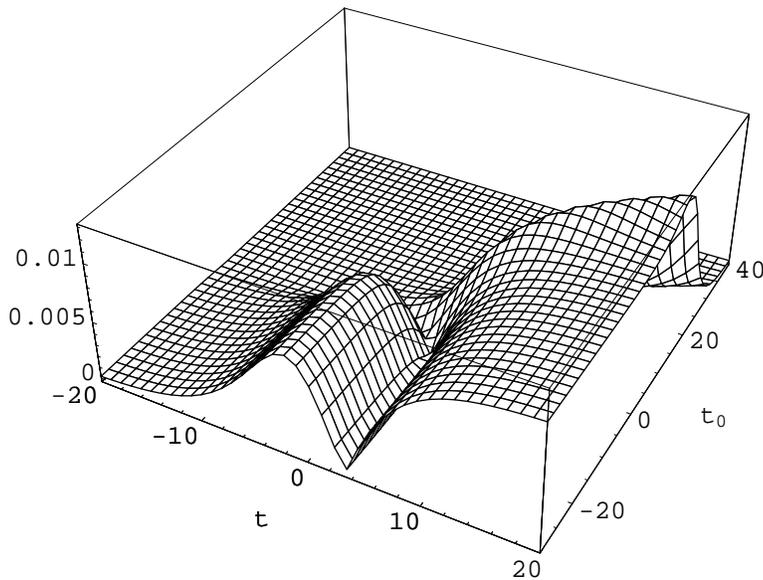

FIG. 6. Morphogenesis of complementarity. The right-hand-side of the uncertainty relation for standard deviations $\Delta P$ and $\Delta P_1$ as a function of time $t$ and the parameter $t_0$ ($t_1 = 0$). Propositions $P$ and $P_1$ are the more complementary the greater the value of this function.

[1] R. Thom, *Structural Stability and Morphogenesis: An Outline of a General Theory of Models* (Benjamin, Reading, 1975).
[2] E. G. Beltrametti, G. Cassinelli, *The Logic of Quantum Mechanics* (Adison-Wesley, Reading, 1981).
[3] D. Aerts, S. Aerts, "Applications of quantum statistics in psychological studies of decision process", Found. Sc. **1**, 85 (1994).
[4] L. Gabora, *Cognitive Mechanisms Underlying the Origin and Evolution of Culture*, Doctoral thesis, CLEA, Brussels Free University (2001).
[5] D. Aerts, B. Coecke, B. D'Hooghe, "A mechanistic macrophysical physical entity with a three dimensional Hilbert space quantum description", Helv. Phys. Acta **70**, 793 (1997).
[6] D. Aerts, S. Aerts, J. Broekaert, L. Gabora, "The violation of Bell inequalities in the macroworld", Found. Phys. **30**, 1387 (2000).
[7] E. W. Piotrowski, J. Sładkowski, "Quantum-like approach to financial risk: Quantum antropic principle", Acta Phys. Polon. B **32**, 3873 (2001).
[8] D. Aerts, "Classical theories and nonclassical theories as a special case of a more general theory", J. Math. Phys. **24**, 2441 (1983).
[9] V. P. Belavkin, V. P. Maslov, "Uniformization method in the theory of nonlinear Hamiltonian Vlasov and Hartree type systems", Theor. Math. Phys. **33**, 17 (1977).
[10] S. Gheorghiu-Svirschevski, "Nonlinear quantum evolution with maximal entropy production", Phys. Rev. A **63**, 022105 (2001).
[11] J. Sładkowski, "Giffen paradoxes in quantum market games", cond-mat/0211083.
[12] L. Glass, M. C. Mackey, *From Clocks to Chaos. The Rythms of Life* (Princeton University Press, Princeton, 1988).
[13] J. D. Murray, *Nonlinear Differential Equation Models in Biology* (Clarendon, Oxford, 1977).
[14] P. C. Fife, *Mathematical Aspects of Reacting and Diffusing Systems*, Lecture Notes in Biomathematics 28 (Springer, New York, 1979).
[15] J. Swift, P. C. Hohenberg, "Hydrodynamic fluctuations at the convective instability", Phys. Rev. A **15**, 319 (1977).
[16] L. N. Howard, N. Kopell, "Slowly varying waves and shock structures in reaction-diffusion equations", Stud. Appl. Math. **56**, 95 (1977).
[17] V. L. Ginzburg, L. D. Landau, "On the theory of superconductivity", Zh. Eksp. Teor. Fiz. **20**, 1064 (1950).
[18] K. Stewartson, J. T. Stuart, "A nonlinear instability theory for a wave system in plane Poiseuille flow", J. Fluid Mech. **48**, 529 (1971).
[19] A. Gierer, H. Meinhardt, "A theory of biological pattern formation", Kybernetik **12**, 30 (1972).

[20] W. Kemmner, "A model of head regeneration in hydra", Differentiation **26**, 83 (1984).
[21] M. Czachor, "Nambu type generalization of the Dirac equation", Phys. Lett. A **225**, 1 (1997)
[22] M. Czachor and J. Naudts, "Microscopic foundation of nonextensive statistics", Phys. Rev. E **59**, R2497 (1999)
[23] S. B. Leble and M. Czachor, "Darboux-integrable nonlinear Liouville-von Neumann equation", Phys. Rev. E **58**, 7091 (1998)
[24] M. Czachor, H. D. Doebner, M. Syty, K. Wasylka, "Von Neumann equations with time-dependent Hamiltonians and supersymetric quantum meachanics", Phys. Rev. E **61**, 3325 (2000).
[25] N. V. Ustinov, M. Czachor, M. Kuna, S. B. Leble, "Darboux integration of $i\dot\rho = [H, f(\rho)]$", Phys. Lett. A **279**, 333 (2001).
[26] A. Peres, "Separability condition for density matrices", Phys. Rev. Lett. **77**, 1413 (1996).
[27] M. Horodecki, P. Horodecki, R. Horodecki, "Separability of mixed states: Necessary and sufficient conditions", Phys. Lett. A **223**, 1 (1996).
[28] D. M. Greenberger, M. A. Horne, A. Shimony, A. Zeilinger, "Bell's theorem without inequalities", Am. J. Phys. **58**, 1131 (1990).
[29] D. Aerts, "A possible explanation for the probabilities of quantum mechanics", J. Math. Phys. **27**, 202 (1986)
[30] S. Bergia, F. Cannata, A. Cornia, R. Livi, "On the actual measurability of the density matrix of a decaying system by means of measurements on the decay products", Found. Phys. **10**, 723 (1980).
[31] N. D. Mermin, "What is quantum mechanics trying to tell us?", Am. J. Phys. **66**, 753 (1998).
[32] C. Walter, R. Parker, M. Ycas, "A model for binary logic in biochemical systems", J. Theor. Biol. **15**, 208 (1967).
[33] L. Glass, S. A. Kauffman, "Co-operative components, spatial localization and oscillatory cellular dynamics", J. Theor. Biol. **34**, 219 (1972).
[34] L. Glass, S. A. Kauffman, "The logical analysis of continuous non-linear biochemical control networks", J. Theor. Biol. **39**, 103(1973).
[35] J. J. Hopfield, D. W. Tank, "Computing with neural circuits: A model", Science **233**, 625 (1986).
[36] C. Tsallis, R. S. Mendes, A. R. Plastino, "The role of constraints within generalized nonextensive statistics", Physica A **261**543 (1998).
[37] S. A. Kauffman, *The Origins of Order. Self-Organization and Selection in Evolution* (Oxford University Press, Oxford, 1993).
[38] E. C. Zeeman, "Catastrophe theory in brain modelling", Int. J. Neurosci. **6**, 39 (1973).
[39] R. Thom, "Topological models in biology", in *Towards a Theoretical Biology*, vol. 3, ed. C. H. Waddington (Aldine, Chicago, 1970).